\documentclass[12pt,twoside]{article}

\usepackage{amssymb}
\usepackage{amsxtra}
\usepackage{amsmath}
\usepackage{amsfonts}
\usepackage{CJK}
\usepackage[titletoc]{appendix}
\usepackage{graphicx}

\numberwithin{equation}{section}

\newtheorem{thm}{Theorem}[section]

\newtheorem{prop}[thm]{Proposition}

\newtheorem{ass}[thm]{Assumption}

\newtheorem{lem}[thm]{Lemma}

\newtheorem{rem}[thm]{Remark}

\newcommand{\eqa}{\begin{eqnarray}}
\newcommand{\eeqa}{\end{eqnarray}}
\newcommand{\beq}{\begin{equation}}
\newcommand{\eeq}{\end{equation}}
\newcommand{\nn}{\nonumber}

\allowdisplaybreaks

\begin{document}

\date{}
\author{Beibei Hu$^{1,2}$ and Tiecheng Xia$^{1,}$\thanks{Corresponding author. E-mails: xiatc@shu.edu.cn(T.-c. Xia), hu\_chzu@163.com(B.-b. Hu) }\\
\small \textit{1.Department of Mathematics, Shanghai University, Shanghai 200444, China}\\
\small \textit{2.School of Mathematical and Finance, Chuzhou University, Anhui 239000, China}}
\title{The coupled modified nonlinear Schr\"{o}dinger equations on the half-line via the Fokas method
\thanks{The work was partially supported by the National
 Natural Science Foundation of China under Grant Nos. 11271008, 61072147, 11601055.}
}
\maketitle
 \hrulefill

\begin{abstract}

Coupled modified nonlinear Schr\"{o}dinger(CMNLS) equations describe the pulse propagation in the picosecond or femtosecond regime of the birefringent optical fibers. In this paper, we use the Fokas method to analyze the initial-boundary value problem for the CMNLS equations on the half-line. Assume that the solution $u(x,t)$ and $v(x,t)$ of CMNLS equations are exists, and we show that it can be expressed in terms of the unique solution of a matrix Riemann-Hilbert problem formulated in the plane of the complex spectral parameter $\lambda$.
\end{abstract}
\hrulefill

\section{Introduction}

\quad\;\;Most important partial differential equations(PDEs) in mathematics and physics
are integrable which can be analysed by the inverse scattering transform(IST) method.
Until the 1990s, the IST method almost can be used to analyze pure initial value
problems. Moreover, in the real world and in some experimental environments, we
need to consider not only the initial conditions but also the boundary conditions.
Therefore, some people began to study the initial boundary value(IBV) problem and
it becomes more interested than the pure initial value problem.

In 1997, Fokas used IST thought to construct a new unified method, we call this method as Fokas method. He analyzed the IBV problems for linear and nonlinear integrable PDEs [1-3].  In the past 20 years, the unified method has been used to analyse boundary value problems for many classical integrable equations with $2\times2$ matrix Lax pairs, such as the Korteweg-deVries(KdV) equation, the nonlinear Schr\"{o}dinger(NLS) equation, the sine-Gordon(sG) equation [4-6], etc. Just like the IST on the line, the unified method provides an expression for the solution of an IBV problem in terms of the solution of a Riemann-Hilbert problem. And the method of Riemann-Hilbert boundary value problem can be used to solve the low friction problem of one-dimensional and three-dimensional quasicrystals in the quasicrystal material fields [7]. In particular, by analyzing the asymptotic behaviour of the solution based on this Riemann-Hilbert problem and by employing the nonlinear version of the steepest descent method introduced by Deift and Zhou [8].

Among those, Fokas and Lenells [9,10], Xu and Fan [11-13] have made a great contribution. In 2012, Lenells [9] applied the unified transform method to analyse IBV problems for integrable evolution equations whose Lax pairs involving $3\times3$ matrices. Following this method, the IBV problems for the Degasperis-Procesi equation was to be studied in [10]. In 2013, Xu and Fan [11] used this method to analyze IBV problem for the Sasa-Satsuma equation, Until 2014, Xu and Fan [12] prove the existence and uniqueness of Riemann-Hilbert problems with $3\times3$ matrix Lax pairs when they analyze the IBV problems for the three wave equation.
After that, more and more researchers begin to pay attention to studying IBV problems for integrable evolution equations with higher order Lax pairs, such as, Geng and Liu [14,15] used this method to study IBV problem for the vector modified KdV equation and the coupled NLS equation on the half-line. Tian [16] used this method to analyze IBV problem for the general coupled NLS equation on the interval, we also have a good time to study partial differential equations with IBV problem.

We consider the following the coupled modified NLS equation in the dimensionless form
\eqa \left\{\begin{array}{l}
iu_t+u_{xx}+\delta[(|u|^2+|v|^2)u]+i\gamma[(|u|^2+|v|^2)u]_x=0,\\
iv_t+v_{xx}+\delta[(|u|^2+|v|^2)v]+i\gamma[(|u|^2+|v|^2)v]_x=0.
\end{array}\right.\label{slisp}\eeqa
where $u=u(x,t)$ and $v=v(x,t)$ is the slowly varying complex envelope for polarizations,
$x$ and $t$ appended to $u,v$ denote partial differentiations,
the parameters $\delta$ and $\gamma$ as real constants are respectively the measure
of cubic nonlinear strength and derivative cubic nonlinearity.
The Eqs.(1.1) has been derived as a model for describe
the propagation of short pulses in birefringent optical fibers both in picosecond and femtosecond
regions [17,18], and which is a hybrid of the coupled NLS equation and coupled derivative NLS equation,
Because case $\gamma=0$ the Eqs.(1.1)) is the known
as the Manakov system [13,14,19], and $\delta=0$ the Eqs.(1.1)
is the coupled derivative NLS equation [20,21].
And some properties of Eqs.(1.1) have been analyzed.
From the integrable viewpoint for nonlinear evolution
equations (NLEEs), Eqs.(1.1) have Lax pairs [18,22],
bilinear representations [22,24], infinitely many
local conservation laws [22].
and the exact bright N-soliton, dark and anti-dark soliton solutions have been
obtained by means of Hirota's bilinear transformation
method [22-25]. In addition, Zhang [26]
have presented the bright vector N-soliton solution by employing Darboux transforma method.
Recentiy, the IBV problem of the system (1.1) with $\gamma=0$
was also obtained quite recently by means of the Fokas method [13,14].

The purpose of this Letter is to analyse the IBV problem of the system (1.1) when ($\delta\neq0,\gamma\neq0$), and the initial boundary values data are defined as
\eqa\begin{array}{l}
u_0(x)=u(x,0),\quad v_0(x)=v(x,0);\\
g_0(t)=u(0,t),\quad h_0(t)=v(0,t);\\
g_1(t)=u_x(0,t),\quad h_1(t)=v_x(0,t).
\end{array}\label{slisp}\eeqa

In this paper, we use the unified transform method to deal with this problem on the half-line domain $\Omega=\{0<x<\infty,0<t<T\}$. We assume that the solution $\{u(x,t),v(x,t)\}$ of Eq.(1.1) are exists. Through this method, we show that it can be expressed in terms of the unique solution of a matrix Riemann-Hilbert problem formulated in the plane of the complex spectral parameter $\lambda$. In addition, we can obtain some spectral functions satisfying the so-called global relation.

The structure of this paper will be arranged as follows. In the next section, we define two sets of eigenfunctions $\mu_j(j=1,2,3)$ and $M_n(n=1,2,3,4)$ of Lax pair for spectral analysis. In the last section, we show that $u(x, t),v(x, t)$ can be expressed in terms of the unique solution of a matrix Riemann-Hilbert problem.

\section{ The spectral analysis}

\quad\;\;We consider the following Lax pair of equations (1.2)[22]
 \eqa \left\{\begin{array}{l}
\psi_x=U\psi=[-\frac{i}{\gamma}(\lambda^2+\frac{\delta}{2})\Lambda+\lambda U_1]\psi,\\
\psi_t=V\psi=[\frac{2i}{\gamma^2}(\lambda^2+\frac{\delta}{2})^2\Lambda-\frac{2}{\gamma}\lambda^3U_1+i\lambda^2U_2-\lambda U_3]\psi,
\end{array}\right. \label{slisp}\eeqa
where $\lambda$ is a complex spectral parameter and
\eqa\begin{array}{l}
\Lambda=\left(\begin{array}{ccc}
-1&0&0\\
0&1&0\\
0&0&1\end{array} \right),\quad\quad\quad
U_1=\left(\begin{array}{ccc}
0& u& v\\
u^*&0&0\\
v^*&0&0\end{array} \right),\\
U_2=\left(\begin{array}{ccc}
-(|u|^2+|v|^2)& 0& 0\\
0&|u|^2&u^*v\\
0&uv^*&|v|^2\end{array} \right),
U_3=\left(\begin{array}{ccc}
0&A_1&A_2\\
A_3&0&0\\
A_4&0&0\end{array} \right),
\end{array}\label{slisp}\eeqa
with
\eqa\begin{array}{l}
A_1=\frac{\delta}{\gamma}u+\gamma u(|u|^2+|v|^2)-iu_x,\quad\quad
A_2=\frac{\delta}{\gamma}v+\gamma v(|u|^2+|v|^2)-iv_x,\nn\\
A_3=\frac{\delta}{\gamma}u^*+\gamma u^*(|u|^2+|v|^2)+iu_x^*,\quad
A_4=\frac{\delta}{\gamma}v^*+\gamma v^*(|u|^2+|v|^2)+iv_x^*.
\end{array}\nn\eeqa

In the following, we let $\delta=2, \gamma=1$ for the convenient of the analysis.

\subsection{The closed one-form}

\quad\;\;We find that Eq.(2.1) is equivalent to
\eqa\begin{array}{l}
\psi_x+ik\Lambda\psi=V_1\psi,\\
\psi_t-2ik^2\Lambda\psi=V_2\psi,
\end{array}\label{slisp}\eeqa
where
 \beq k=\lambda^2+1,\quad V_1=\lambda U_1,\quad V_2=-2\lambda^3U_1+i\lambda^2U_2-\lambda U_3,\quad\delta=2, \gamma=1.\label{slisp}\eeq

We assume that $u(x,t),v(x,t)$ is a sufficiently smooth function in the half-line region $\Omega=\{0<x<\infty,0<t<T\}$, and decays sufficiently when $x\rightarrow\infty.$ Introducing a new function $\mu(x,t,\lambda)$ by
\beq\psi=\mu e^{-ik\Lambda x+2ik^2\Lambda t},\label{slisp}\eeq
and the corresponding Lax pair equation Eq.(2.3) becomes
\eqa\begin{array}{l}
\mu_x+ik[\Lambda,\mu]=V_1\mu,\\
\mu_t-2ik^2[\Lambda,\mu]=V_2\mu,
\end{array}\label{slisp}\eeqa
then Eq.(2.6) can be written to the differential form
\beq d(e^{ik\hat\Lambda x-2ik^2\hat\Lambda t}\mu)=W(x,t,\lambda), \label{slisp}\eeq
where $W(x,t,\lambda)$ can be defined as
\beq W(x,t,\lambda)=e^{(ik x-2ik^2t)\hat\Lambda\mu}(V_1dx+V_2dt)\mu, \label{slisp}\eeq
and $\hat\Lambda$ represents a matrix operator acting on B by $\hat\Lambda B=[\Lambda,B]$.

\subsection{ The eigenfunction}

\quad\;\;There are three eigenfunctions $\mu_j(x,t,\lambda)(j=1,2,3)$ of Eq.(2.6) which are defined by the following the Volterra integral equation
\beq \mu_j(x,t,\lambda)=\mathbb{I}+\int_{\gamma_j}e^{(-ik x+2ik^2t)\hat\Lambda}W_j(x,t,\lambda),\quad j=1,2,3,\label{slisp}\eeq
where $W_j$ is given by Eq.(2.8), it is only used $\mu_j$ in place of $\mu$, and the contours $ \gamma_j(j=1,2,3)$ are shown in figure 1.

\begin{figure}[ht]
\centering
\includegraphics[width=5.4in,height=2.5in]{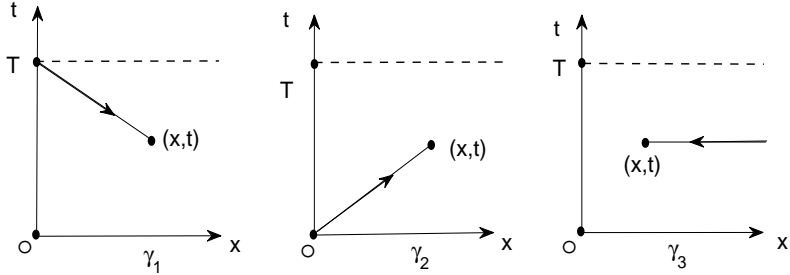}
\caption{The three contours $\gamma_1,\gamma_2,\gamma_3$ in the $(x,t)$-domaint}\label{fig:graph}
\end{figure}

The first, second, and third columns of the matrix equation (2.9) contain the following exponential term
\eqa\begin{array}{l}
\mu_j^{(1)}:\quad e^{-2ik(x-\xi)+4ik^2(t-\tau)}, e^{-2ik(x-\xi)+4ik^2(t-\tau)};\\
\mu_j^{(2)}:\quad e^{2ik(x-\xi)-4ik^2(t-\tau)};\\
\mu_j^{(3)}:\quad e^{2ik(x-\xi)-4ik^2(t-\tau)}.
\end{array}\label{slisp}\eeqa
At the same time, the following inequalities hold true on the contours
\eqa \begin{array}{l}
\gamma_1:\quad x-\xi\geq 0,\quad t-\tau\leq 0;\\
\gamma_2:\quad x-\xi\geq 0,\quad t-\tau\geq 0;\\
\gamma_3:\quad x-\xi\leq 0.\\
\end{array}\label{slisp}\eeqa
Thus, these inequalities imply that the eigenfunctions $\mu_j(x,t,\lambda)(j=1,2,3)$ are bounded and analytic for $\lambda\in\mathbb{C}$ such that $\lambda$ belongs to
\eqa\begin{array}{l}
 \mu_1 $ is bounded and analytic for $ \lambda\in(D_4,D_1,D_1),\\
 \mu_2 $ is bounded and analytic for $ \lambda\in(D_3,D_2,D_2),\\
 \mu_3 $ is bounded and analytic for $ \lambda\in(D_1\cup D_2,D_3\cup D_4,D_3\cup D_4),\\
\end{array}\label{slisp}\eeqa
where $D_n(n=1,2,3,4)$ represents a subset of four open disjoint $\lambda$-plane shown in figure 2.

\begin{figure}[ht]
\centering
\includegraphics[width=7.4in,height=3.0in]{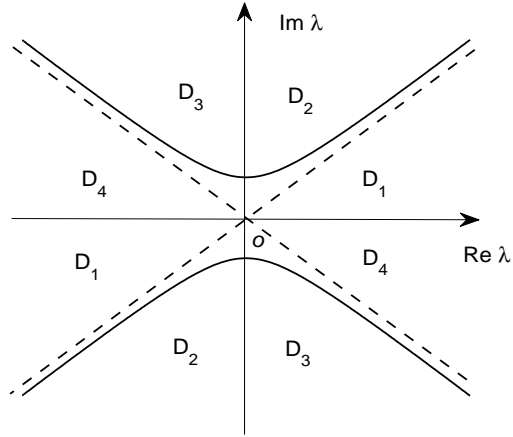}
\caption{The sets $D_j,j=1,2,3,4$, which decompose the complex $\lambda-$plane}\label{fig:graph}
\end{figure}

And these sets $D_n(n=1,2,3,4)$ have the following properties
\eqa\begin{array}{l}
D_1=\{\lambda\in\mathbb{C}|Rel_1<Rel_2=Rel_3,\quad Rez_1>Rez_2=Rez_3\},\\
D_2=\{\lambda\in\mathbb{C}|Rel_1<Rel_2=Rel_3,\quad Rez_1<Rez_2=Rez_3\},\\
D_3=\{\lambda\in\mathbb{C}|Rel_1>Rel_2=Rel_3,\quad Rez_1>Rez_2=Rez_3\},\\
D_4=\{\lambda\in\mathbb{C}|Rel_1>Rel_2=Rel_3,\quad Rez_1<Rez_2=Rez_3\},
\end{array}\label{slisp}\eeqa
where $l_i(\lambda)$ and $z_i(\lambda)$ are the diagonal elements of the matrix $-ik\Lambda$ and $2ik^2\Lambda$.

In fact, $\mu_1(0,t,\lambda)$ has a larger bounded and analytic domain is $(D_1\cup D_3,D_2\cup D_4,D_2\cup D_4)$ for $x=0$, and $\mu_2(0,t,\lambda)$ also has a larger bounded and analytic domain is $(D_2\cup D_4,D_1\cup D_3,D_1\cup D_3)$ for $x=0$.

For each $n=1,2,3,4$, the solution $M_n(x,t,\lambda)$ of Eq.(2.6) is defined by the following integral equation
\beq (M_n(x,t,\lambda))_{ij}=\delta_{ij}+\int_{\gamma_{ij}^n}(e^{(-ik x+2ik^2t)\hat\sigma}W_n(\xi,\tau,\lambda))_{ij},
\quad i,j=1,2,3,\label{slisp}\eeq
where $W_n(x,t,\lambda)$ is given by Eq.(2.8), it is only used $M_n$ in place of $\mu$, and the contours $ \gamma_{ij}^n (n=1,2,3,4;i,j=1,2,3)$ are defined as follows
 \eqa \gamma_{ij}^n=\left\{\begin{array}{l}
 \gamma_1\quad if\quad Rel_i(\lambda)<Rel_j(\lambda) \quad and \quad Rez_i(\lambda)\geq Rez_j(\lambda),\\
 \gamma_2\quad if\quad Rel_i(\lambda)<Rel_j(\lambda) \quad and \quad Rez_i(\lambda)<Rez_j(\lambda),\quad for\lambda\in D_n\\
 \gamma_3\quad if \quad Rel_i(\lambda)\geq Rel_j(\lambda).
\end{array}\right. \label{slisp}\eeqa
According to the definition of $\gamma^n$, we have
\eqa \begin{array}{l}
\gamma^1=\left(\begin{array}{ccc}
\gamma_3&\gamma_1&\gamma_1\\
\gamma_3&\gamma_3&\gamma_3\\
\gamma_3&\gamma_3&\gamma_3\end{array} \right),\quad
\gamma^2=\left(\begin{array}{ccc}
\gamma_3&\gamma_2&\gamma_2\\
\gamma_3&\gamma_3&\gamma_3\\
\gamma_3&\gamma_3&\gamma_3\end{array} \right),\\
\gamma^3=\left(\begin{array}{ccc}
\gamma_3&\gamma_3&\gamma_3\\
\gamma_2&\gamma_3&\gamma_3\\
\gamma_2&\gamma_3&\gamma_3\end{array} \right),\quad
\gamma^4=\left(\begin{array}{ccc}
\gamma_3&\gamma_3&\gamma_3\\
\gamma_1&\gamma_3&\gamma_3\\
\gamma_1&\gamma_3&\gamma_3\end{array} \right).
\end{array}\label{slisp} \eeqa

Next, the following proposition guarantees that the previous definition of $M_n$ has properties, namely, $M_n$ can be represented as a Rimann-Hilbert problem.

\begin{prop} For each $n=1,2,3,4$ and $\lambda\in D_n$, the function $M_n(x,t,\lambda)$ is defined well by Eq.(2.14). For any identified point $(x,t)$, $M_n$
is bounded and analytical as a function of $\lambda\in D_n$ away from a possible discrete set of singularities $\{\lambda_j\}$ at which the Fredholm determinant vanishes. Moreover, $M_n$ admits a bounded and continuous extension to $\bar D_n$ and
\beq M_n(x,t,\lambda)=\mathbb{I}+\mathcal{O}(\frac{1}{\lambda}).\label{slisp}\eeq\end{prop}

Proof: The associated bounded and analytical properties have been established in Appendix B in \cite{Lenells2012}. Substituting the following expansion
\beq M=M_0+\frac{M^{(1)}}{\lambda}+\frac{M^{(2)}}{\lambda^2}+\cdots\quad\quad\lambda\rightarrow\infty,\nn\eeq
into the Lax pair Eq.(2.6) and comparing the coefficients of the same order of $\lambda$, we can obtain Eq.(2.17).
\begin{rem}So far, we define two sets of eigenfunctions $\mu_j(j=1,2,3)$ and $M_n(n=1,2,3,4)$. The Fokas method in [1] analyzed
 the $2\times2$ Lax pair related to two kinds of eigenfunctions $\mu_j$, which is used for spectral analysis, and the other eigenfunction is used to be shown Riemann-Hilbert problem, our definition on $M_n$ is similar to the latter eigenfunction.\end{rem}

\subsection{ The jump matrix}

\quad\;\;We define the spectral function as follows
\beq S_n(\lambda)=M_n(0,0,\lambda),\quad \lambda\in D_n,n=1,2,3,4.\label{slisp}\eeq
Let $M$ be a sectionally analytical continuous function in Riemann $\lambda$-sphere which equals $M_n$ for
$\lambda\in D_n$. Then $M$ satisfies the following jump conditions
\beq M_n(\lambda)=M_mJ_{m,n},\quad \lambda\in \bar D_n\cap \bar D_m,\quad n,m=1,2,3,4;n\neq m,\label{slisp}\eeq
where $J_{m,n}=J_{m,n}(x,t;\lambda)$ are the jump matrices given by
\beq J_{m,n}=e^{(-ik x+2ik^2t)\hat\Lambda}(S_m^{-1}S_n).\label{slisp}\eeq

\begin{rem}As the integral equation (2.14) defined by $M_n(0,0,\lambda)$ involves only along the initial half-line $\{0<x<\infty,t=0\}$
and along the boundary $\{x=0,0<t<T\}$ , so $S_n$'s(and hence also the $J_{m,n}$'s) can only be determined by the initial data
and boundary data, therefore, equation (2.20) represents a jump condition of Riemann-Hilbert problem.
In the absence of singularity, the solution $u(x,t),v(x,t)$ of the equation can be reconstructed from the initial data and boundary values data,
but if the $M_n$ have pole singularities at some point $\{\lambda_j\},\lambda_j\in\mathbb{C}$, the Riemann-Hilbert problem
should be included the residue condition in these points, so in order to determine the correct residue condition,
we need to introduce three eigenfunctions $\mu_j(x,t,\lambda)(j=1,2,3)$ in addition to the $M_n$'s.\end{rem}

\subsection{ The adjugated eigenfunction}

\quad\;\;We also need to consider the bounded and analytical properties of the minors of the matrices $\mu_j(x,t,\lambda)(j=1,2,3)$. We recall that the cofactor matrix $B^A$ of a $3\times3$ matrix $B$ is defined by
$$ B^A=\left(\begin{array}{ccc}
m_{11}(B)&-m_{12}(B)&m_{13}(B)\\
-m_{21}(B)&m_{22}(B)&-m_{23}(B)\\
m_{31}(B)&-m_{32}(B)&m_{33}(B) \end{array}
\right),$$
where $m_{ij}(B)$ denote the $(ij)$th minor of $B$. From Eq.(2.6) we find that the adjugated eigenfunction $\mu^A$ satisfies the Lax pair
\eqa\begin{array}{l}
\mu_x^A-ik[\Lambda,\mu^A]=-V_1^T\mu^A,\\
\mu_t^A+2ik^2[\Lambda,\mu^A]=-V_2^T\mu^A,\\
\end{array}\label{slisp}\eeqa
where the superscript $T$ denotes a matrix transpose. Then the eigenfunctions $\mu_j(j=1,2,3)$ are solutions of the integral equations
\beq \mu_j^A(x,t,\lambda)=\mathbb{I}-\int_{\gamma_j}e^{(ik(x-\xi)-2ik^2(t-\tau))\hat\Lambda}(V_1^Tdx+V_2^Tdt),\quad j=1,2,3.\label{slisp}\eeq
Thus, we can obtain the adjugated eigenfunction which satisfies the following analytic properties
\eqa\begin{array}{l}
 \mu_1^A $ is bounded and analytic for $ \lambda\in(D_1,D_4,D_4),\\
 \mu_2^A $ is bounded and analytic for $ \lambda\in(D_2,D_3,D_3),\\
 \mu_3^A $ is bounded and analytic for $ \lambda\in(D_3\cup D_4,D_1\cup D_2,D_1\cup D_2).
\end{array}\label{slisp}\eeqa
In fact, $\mu_1^A(0,t,\lambda)$ has a larger bounded and analytic domain which is $(D_2\cup D_4,D_1\cup D_3,D_1\cup D_3)$ for $x=0$, and $\mu_2^A(0,t,\lambda)$ also has a larger bounded analytic domain which is $(D_1\cup D_3,D_2\cup D_4,D_2\cup D_4)$ for $x=0$.

\subsection{ Symmetry}

\quad\;\;By the following Lemma, we show that the eigenfunctions $\mu_j(x,t,\lambda)$ have an important symmetry.
\begin{lem} The eigenfunction $\psi(x,t,\lambda)$ of the Lax pair Eq.(2.1) admits the following symmetry
$$\psi^{-1}(x,t,\lambda)=A \overline{\psi(x,t,\bar\lambda)}^TA,$$
with
$$
A=\left(\begin{array}{ccc}
-1&0&0\\
0&\varepsilon&0\\
0&0&\varepsilon\end{array}
\right),\quad and \quad\varepsilon^2=1.$$
where the superscript T denotes a matrix transpose.\end{lem}

Proof: Analogous to the proof provided in [9]. We omit the proof.
\begin{rem} From Lemma 2.4, one can show that the eigenfunctions $\mu_j(x,t,\lambda)$ of Lax pair Eq.(2.6) admit the same symmetry.\end{rem}

\subsection{ The jump matrix computations}

\quad\;\;We also define the $3\times3$ matrix value spectral function $s(\lambda)$ and $S(\lambda)$ as follows
\eqa\begin{array}{l}
\mu_3(x,t,\lambda)=\mu_2(x,t,\lambda)e^{(-ik x+2ik^2t)\hat\Lambda} s(\lambda),\\
\mu_1(x,t,\lambda)=\mu_2(x,t,\lambda)e^{(-ik x+2ik^2t)\hat\Lambda} S(\lambda),\\
\end{array}\label{slisp}\eeqa
as $\mu_2(0,0,\lambda)=\mathbb{I}$, we obtain
\beq s(\lambda)=\mu_3(0,0,\lambda),\quad\quad S(\lambda)=\mu_1(0,0,\lambda).\label{slisp}\eeq

From the properties of $\mu_j$ and $\mu_j^A(j=1,2,3)$  we can obtain that $s(\lambda)$ and $S(\lambda)$ have the following bounded and analytic properties
\eqa\begin{array}{l} s(\lambda) $ is bounded for $ \lambda\in(D_1\cup D_2,D_3\cup D_4,D_3\cup D_4),\\
 S(\lambda)$ is bounded for $ \lambda\in(D_1\cup D_3,D_2\cup D_4,D_2\cup D_4),\\
 s^A(\lambda)$ is bounded for $ \lambda\in(D_3\cup D_4,D_1\cup D_2,D_1\cup D_2),\\
 S^A(\lambda)$ is bounded for $ \lambda\in(D_2\cup D_4,D_1\cup D_3,D_1\cup D_3).
\end{array}\label{slisp}\eeqa
Moreover
\beq M_n(x,t,\lambda)=\mu_2(x,t,\lambda)e^{(-ik x+2ik^2t)\hat\Lambda} S_n(\lambda),\quad \lambda\in D_n.\label{slisp}\eeq
\begin{prop}
The $S_n$ can be expressed with $s(\lambda)$ and $S(\lambda)$  elements as follows
\eqa \begin{array}{l}  S_1=\left(\begin{array}{ccc}
s_{11}&\frac{m_{33}(s)M_{21}(S)-m_{23}(s)M_{31}(S)}{(s^TS^A)_{11}}&\frac{m_{32}(s)M_{21}(S)-m_{22}(s)M_{31}(S)}{(s^TS^A)_{11}}\\
s_{21}&\frac{m_{33}(s)M_{11}(S)-m_{13}(s)M_{31}(S)}{(s^TS^A)_{11}}&\frac{m_{32}(s)M_{11}(S)-m_{12}(s)M_{31}(S)}{(s^TS^A)_{11}}\\
s_{31}&\frac{m_{23}(s)M_{11}(S)-m_{13}(s)M_{21}(S)}{(s^TS^A)_{11}}&\frac{m_{22}(s)M_{11}(S)-m_{12}(s)M_{21}(S)}{(s^TS^A)_{11}}
\end{array} \right),\\
S_2=\left(\begin{array}{ccc}
s_{11}&0&0\\
s_{21}&\frac{m_{33}(s)}{s_{11}}&\frac{m_{32}(s)}{s_{11}}\\
s_{31}&\frac{m_{23}(s)}{s_{11}}&\frac{m_{22}(s)}{s_{11}}
\end{array} \right),\quad
S_3=\left(\begin{array}{ccc}
\frac{1}{m_{11}(s)}&s_{12}&s_{13}\\
0&s_{22}&s_{23}\\
0&s_{32}&s_{33}
\end{array} \right),\\
S_4=\left(\begin{array}{ccc}
\frac{S_{11}}{(S^Ts^A)_{11}}&s_{12}&s_{13}\\
\frac{S_{21}}{(S^Ts^A)_{11}}&s_{22}&s_{23}\\
\frac{S_{31}}{(S^Ts^A)_{11}}&s_{32}&s_{33}
\end{array} \right).\end{array}\label{slisp}\eeqa
\end{prop}

Proof: We set that $\gamma_3^{X_0}$ is a contour when $(X_0,0)\rightarrow (x,t)$ in the $(x,t)$-plane, here $X_0$ is a constant and $X_0>0$, for $j=3$, we introduce $\mu_3(x,t,\lambda;X_0)$ as the solution of Eq.(2.9) with the contour $\gamma_3$ replaced by $\gamma_3^{X_0}$. Similarly, we define $M_n(x,t,\lambda;X_0)$ as the solution of Eq.(2.14) with $\gamma_3$ replaced by $\gamma_3^{X_0}$. then, by simple calculation, we can use $S(\lambda)$ and $s(\lambda;X_0)=\mu_3(0,0,\lambda;X_0)$ to derive the expression of $S_n(\lambda,X_0)=M_n(0,0,\lambda;X_0)$ and the Eq.(2.28) will be obtained by taking the limit $X_0\rightarrow\infty$.

Firstly, we have the following relations:
\beq M_n(x,t,\lambda;X_0)=\mu_1(x,t,\lambda)e^{(-ik x+2ik^2t)\hat\Lambda} R_n(\lambda;X_0),\label{slisp}\eeq
\beq M_n(x,t,\lambda;X_0)=\mu_2(x,t,\lambda)e^{(-ik x+2ik^2t)\hat\Lambda} S_n(\lambda;X_0),\label{slisp}\eeq
\beq M_n(x,t,\lambda;X_0)=\mu_3(x,t,\lambda)e^{(-ik x+2ik^2t)\hat\Lambda} T_n(\lambda;X_0).\label{slisp}\eeq

Secondly, we can get the definition of $R_n(\lambda;X_0)$ and $T_n(\lambda;X_0)$ as follows
\beq R_n(\lambda;X_0)=e^{-2ik^2T\hat\Lambda} M_n(0,T,\lambda;X_0),\label{slisp}\eeq
\beq T_n(\lambda;X_0)=e^{ik X_0\hat\Lambda} M_n(X_0,0,\lambda;X_0),\label{slisp}\eeq
then equations(2.29),(2.30) and (2.31) mean that
\beq s(\lambda;X_0)=S_n(\lambda;X_0)T_n^{-1}(\lambda;X_0),\label{slisp}\eeq
\beq S(\lambda;X_0)=S_n(\lambda;X_0)R_n^{-1}(\lambda;X_0).\label{slisp}\eeq

These equations constitute the matrix decomposition problem of $\{s,S\}$ by use $\{R_n,S_n,T_n\}$. In fact, by the definition of the integral equation (2.14) and $\{R_n,S_n,T_n\}$, we obtain
{\eqa\left\{\begin{array}{l}
(R_n(\lambda;X_0))_{ij}=0\quad if \quad \gamma_{ij}^n=\gamma_1,\\
(S_n(\lambda;X_0))_{ij}=0\quad if \quad \gamma_{ij}^n=\gamma_2,\\
(T_n(\lambda;X_0))_{ij}=\delta_{ij}\quad if \quad \gamma_{ij}^n=\gamma_3.
\end{array}\right. \label{slisp}\eeqa}

Thus equations (2.34) and (2.35) are the 18 scalar equations with 18 unknowns. The exact solution of these system can be obtained by solving the algebraic system,
 in this way, we can get a similar $\{S_n(\lambda),s(\lambda)\}$ as in Eq.(2.28) which just that $\{S_n(\lambda),s(\lambda)\}$ replaces by $\{S_n(\lambda;X_0),s(\lambda;X_0)\}$ in Eq.(2.28).

 Finally, taking $X_0\rightarrow\infty$ in this equation, we obtain the Eq.(2.28).

\subsection{ The residue conditions}

\quad\;\;Because $\mu_2$ is an entire function, and from Eq.(2.27) we know that $M$ only have singularities at the points where the in $S_n (n=1,2,3,4)$ have singularities. We introduce the symbols $\lambda_j(j=1,2\cdots N)$ to denote the possible zeros, and assume that the $\lambda_j(j=1,2\cdots N)$ satisfy the following assumption.
\begin{ass}We assume that\end{ass}

(1)$(s^TS^A)_{11}(\lambda)$ possess $n_0\geq0$ possible simple zeros in $D_1$ denoted by $\lambda_j,j=1,2\cdots n_0$,

(2)$s_{11}(\lambda)$ possess $n_1-n_0\geq0$ possible simple zeros in $D_2$ denoted by $\lambda_j,j=n_0+1,n_0+2\cdots n_1$,

(3)$m_{11}(s)(\lambda)$ possess $n_2-n_1\geq0$ possible simple zeros in $D_3$ denoted by $\lambda_j,j=n_1+1,n_1+2\cdots n_2$,

(4)$(S^Ts^A)_{11}(\lambda)$ possess $N-n_2\geq0$ possible simple zeros in $D_4$ denoted by $\lambda_j,j=n_2+2,n_2+2\cdots N$,

And these zeros are each different, moreover, we assuming that none of such functions $(s^TS^A)_{11}(\lambda),s_{11}(\lambda),m_{11}(s)(\lambda)$
and $(S^Ts^A)_{11}(\lambda)$ have zeros on the boundaries of the $D_n(n=1,2,3,4)$.

\begin{prop}
Assume that $M_n(n=1,2,3,4)$ are the eigenfunctions defined by (2.14) and the set $\lambda_j(j=1,2\cdots N)$ of singularities are as the above assumption. Then the following residue conditions hold true:
\beq \begin{array}{l} Res_{\lambda=\lambda_j}[M]_2=\frac{m_{33}(s)(\lambda_j)M_{11}(S)(\lambda_j)-m_{13}(s)(\lambda_j)M_{31}(S)(\lambda_j)}{\dot{(s^TS^A)_{11}(\lambda_j)}s_{21}(\lambda_j)}e^{\theta_{13}(\lambda_j)}[M(\lambda_j)]_1
,\\ \quad\quad\quad\quad\quad\quad\quad\quad\quad\quad\quad\quad\quad\quad\quad\quad\quad\quad\quad\quad
 1\leq j\leq n_0;\lambda_j\in D_1.\end{array}\label{slisp}\eeq
\beq \begin{array}{l} Res_{\lambda=\lambda_j}[M]_3=\frac{m_{32}(s)(\lambda_j)M_{11}(S)(\lambda_j)-m_{12}(s)(\lambda_j)M_{31}(S)(\lambda_j)}{\dot{(s^TS^A)_{11}(\lambda_j)}s_{21}(\lambda_j)}e^{\theta_{13}(\lambda_j)}[M(\lambda_j)]_1
,\\ \quad\quad\quad\quad\quad\quad\quad\quad\quad\quad\quad\quad\quad\quad\quad\quad\quad\quad\quad\quad
1\leq j\leq n_0;\lambda_j\in D_1.\end{array}\label{slisp}\eeq
\beq \begin{array}{l}
Res_{\lambda=\lambda_j}[M]_2=\frac{m_{33}(s)(\lambda_j)}{\dot{s_{11}(\lambda_j)}s_{21}(\lambda_j)}e^{\theta_{13}(\lambda_j)}[M(\lambda_j)]_1
,\quad n_0+1\leq j\leq n_1;\lambda_j\in D_2.\end{array}\label{slisp}\eeq
\beq \begin{array}{l}
Res_{\lambda=\lambda_j}[M]_3=\frac{m_{32}(s)(\lambda_j)}{\dot{s_{11}(\lambda_j)}s_{21}(\lambda_j)}e^{\theta_{13}(\lambda_j)}[M(\lambda_j)]_1
,\quad n_0+1\leq j\leq n_1;\lambda_j\in D_2.\end{array}\label{slisp}\eeq
\beq \begin{array}{l} Res_{\lambda=\lambda_j}[M]_1=\frac{s_{33}(\lambda_j)[M_(\lambda_j)]_2-s_{32}(\lambda_j)[M_(\lambda_j)]_3}{\dot{m_{11}(s)(\lambda_j)}m_{21}(s)(\lambda_j)}e^{\theta_{31}(\lambda_j)}
, n_1+1\leq j\leq n_2;\lambda_j\in D_3.\end{array}\label{slisp}\eeq
\beq \begin{array}{l} Res_{\lambda=\lambda_j}[M]_1=\frac{s_{33}(\lambda_j)S_{21}(\lambda_j)-s_{23}(\lambda_j)S_{31}(\lambda_j)}{\dot{(S^Ts^A)_{11}(\lambda_j)}m_{11}(s)(\lambda_j)}e^{\theta_{31}(\lambda_j)}[M(\lambda_j)]_2
\\ \quad\quad\quad+\frac{s_{22}(\lambda_j)S_{31}(\lambda_j)-s_{32}(\lambda_j)S_{21}(\lambda_j)}{\dot{(S^Ts^A)_{11}(\lambda_j)}m_{11}(s)(\lambda_j)}e^{\theta_{31}(\lambda_j)}[M(\lambda_j)]_3
,n_2+1\leq j\leq N;\lambda_j\in D_4.\end{array}\label{slisp}\eeq
where $\dot{f}=\frac{df}{d\lambda}$ and $\theta_{ij}$ defined by
\beq \theta_{ij}(x,t,\lambda)=(l_i-l_j)x-(z_i-z_j)t, \quad i,j=1,2,3;\label{slisp}\eeq
thus
\beq \theta_{ij}=0,\quad i,j=2,3;\quad\quad \theta_{12}=\theta_{13}=-\theta_{21}=-\theta_{31}=2ik x+4ik^2t.\nn\eeq
\end{prop}

Proof: We will only prove (2.37), (2.38) and the other conditions follow by similar arguments. The equation (2.27) mean that
\beq M_1=\mu_2e^{(-ik x+2ik^2t)\hat\Lambda} S_1,\label{slisp}\eeq

In view of the expressions for $S_1$ given in (2.28), the three columns of Eq.(2.44) read
\beq [M_1]_1=[\mu_2]_1s_{11}+[\mu_2]_2s_{21}e^{\theta_{31}}+[\mu_2]_3s_{31}e^{\theta_{31}},\label{slisp}\eeq
\eqa&&
[M_1]_2=\frac{m_{33}(s)M_{21}(S)-m_{23}(s)M_{31}(S)}{(s^TS^A)_{11}}e^{\theta_{13}}[\mu_2]_1+\frac{m_{33}(s)M_{11}(S)-m_{13}(s)M_{31}(S)}{(s^TS^A)_{11}}[\mu_2]_2
\nn\\&&\quad\quad\quad\quad+\frac{m_{23}(s)M_{11}(S)-m_{13}(s)M_{21}(S)}{(s^TS^A)_{11}}[\mu_2]_3,\label{slisp}\eeqa
\eqa&&
[M_1]_3=\frac{m_{32}(s)M_{21}(S)-m_{22}(s)M_{31}(S)}{(s^TS^A)_{11}}e^{\theta_{13}}[\mu_2]_1+\frac{m_{32}(s)M_{11}(S)-m_{12}(s)M_{31}(S)}{(s^TS^A)_{11}}[\mu_2]_2
\nn\\&&\quad\quad\quad\quad+\frac{m_{22}(s)M_{11}(S)-m_{12}(s)M_{21}(S)}{(s^TS^A)_{11}}[\mu_2]_3.\label{slisp}\eeqa

Let $\lambda_j\in D_1$ be a simple zero of $(s^TS^A)_{11}(\lambda)$. Solving Eq.(2.45) for $[\mu_2]_2$ and substituting
the result into Eq.(2.46) and  Eq.(2.47), we find
\beq [M_1]_2=\frac{m_{33}(s)M_{11}(S)-m_{13}(s)M_{31}(S)}{(s^TS^A)_{11}s_{21}}e^{\theta_{13}}[M_1]_1-\frac{m_{33}(s)}{s_{21}}e^{\theta_{13}}[\mu_2]_1
+\frac{m_{13}(s)}{s_{21}}[\mu_2]_3,\label{slisp}\eeq
\beq [M_1]_3=\frac{m_{32}(s)M_{11}(S)-m_{12}(s)M_{31}(S)}{(s^TS^A)_{11}s_{21}}e^{\theta_{13}}[M_1]_1-\frac{m_{32}(s)}{s_{21}}e^{\theta_{13}}[\mu_2]_1
+\frac{m_{12}(s)}{s_{21}}[\mu_2]_3.\label{slisp}\eeq
Taking the residue of the two equation at $\lambda_j$, we find condition Eq.(2.37) and Eq.(2.38) in the case when $\lambda_j\in D_1$.

\subsection{ The global relation }

\quad\;\;The spectral functions $S(\lambda)$ and $s(\lambda)$ are not independent which is of important relationship each other. In fact, from Eq.(2.24), we find
\beq \mu_3(x,t,\lambda)=\mu_1(x,t,\lambda)e^{(-ik x+2ik^2t)\hat\Lambda} S^{-1}(\lambda)s(\lambda),
\lambda\in(D_1\cup D_2,D_3\cup D_4,D_3\cup D_4),\label{slisp}\eeq
as $\mu_1(0,t,\lambda)=\mathbb{I}$, when $(x,t)=(0,T)$, We can evaluate the following relationship which is the global relation as follows
\beq S^{-1}(\lambda)s(\lambda)=e^{-2ik^2T\hat\Lambda}c(T,\lambda),\quad
\lambda\in(D_1\cup D_2,D_3\cup D_4,D_3\cup D_4),
\label{slisp}\eeq
where  $c(T,\lambda)=\mu_3(0,t,\lambda)$.

\section{The Riemann-Hilbert problem }

\quad\;\;In section 2, we define the sectionally analytical function $M(x,t,\lambda)$ that its satisfies a Riemann-Hilbert problem which can be formulated in terms of the initial and boundary values of $\{u(x,t),v(x,t)\}$. For all $(x,t)$, the solution of Eq.(1.1) can be recovered by solving this Riemann-Hilbert problem. So we can establish the following theorem.

\begin{thm}Suppose that $\{u(x,t),v(x,t)\}$ are solution of Eq.(1.1) in the half-line domain $\Omega$, and it is sufficient smoothness and decays when $x\rightarrow\infty$. Then the $\{u(x,t),v(x,t)\}$ can be reconstructed from the initial values $\{u_0(x),v_0(x)\}$ and boundary values $\{g_0(t),h_0(t),g_1(t),h_1(t)\}$ defined as follows
\eqa\begin{array}{l}
u_0(x)=u(x,0),\quad v_0(x)=v(x,0);\\
g_0(t)=u(0,t),\quad h_0(t)=v(0,t);\\
g_1(t)=u_x(0,t),\quad h_1(t)=v_x(0,t).
\end{array}\label{slisp}\eeqa
Like Eq.(2.24) using the initial and boundary data to define the spectral functions $s(\lambda)$ and $S(\lambda)$,we can further define the jump matrix $J_{m,n}(x,t,\lambda)$. Assume that the zero points of the  $s_{11}(\lambda),(s^TS^A)_{11}(\lambda),(S^Ts^A)_{11}(\lambda)$ and $m_{11}(s)(\lambda)$ are $\lambda_j(j=1,2\cdots N)$ just like in assumption 2.7. We also have the following results
\eqa\begin{array}{l}
u(x,t)=2i\lim_{\lambda\rightarrow\infty}(\lambda M(x,t,\lambda))_{12},\\
v(x,t)=2i\lim_{\lambda\rightarrow\infty}(\lambda M(x,t,\lambda))_{13}.
\end{array}\label{slisp}\eeqa
where $M(x,t,\lambda)$ satisfies the following $3\times 3$ matrix Riemann-Hilbert problem:

(1)$M$ is a sectionally meromorphic on the Riemann $\lambda$-sphere with jumps across the contours on $\bar D_n\cap\bar D_m(n,m=1,2,3,4)$ (see figure 2).

(2)$M$ satisfies the jump condition with jumps across the contours on $\bar D_n\cap\bar D_m(n,m=1,2,3,4)$
\beq M_n(\lambda)=M_mJ_{m,n},\quad \lambda\in \bar D_n\cap \bar D_m,n,m=1,2,3,4;n\neq m.\label{slisp}\eeq

(3)$M(x,t,\lambda)=\mathbb{I}+\mathcal{O}(\frac{1}{\lambda}),\quad \lambda\rightarrow\infty.$

(4)The residue condition of $M$ is showed in Proposition 2.8.

\end{thm}

Proof: We can use similar method with [11] to prove this Theorem, It only remains to prove Eq.(3.2) and this equation follows from the large $\lambda$ asymptotic of the eigenfunctions. We omit this proof in here because of the length of this article.

\begin{flushleft}
\textbf{Reference}
\end{flushleft}

[1]A. S. Fokas, A unified transform method for solving linear and certain nonlinear PDEs,
Proc R Soc Lond A. 453:1411-1443(1997).

[2]A. S. Fokas, Integrable nonlinear evolution equations on the half-line, Commun Math Phys.
230:1-39(2002).

[3]A. S. Fokas, A unified approach to boundary value problems, CBMS-NSF Regional Conference
Series in Applied Mathematics, Philadelphia, PA: Society of Industrial and Applied
Mathematics, 2008.

[4]J. Lenells, and A. S. Fokas, Boundary-value problems for the stationary axisymmetric Einstein
equations, a rotating disc, Nonlinearity. 24:177-206(2011).

[5]A. S. Fokas, A. R. Its, and L. Y. Sung, The nonlinear Schr\"{o}dinger equation on the half-line,
Nonlinearity. 18:1771-1822(2005).

[6]J. Lenells, Boundary value problems for the stationary axisymmetric Einstein equations,
a disk rotating around a black hole, Comm Math Phys. 304:585-635(2011).

[7]X. Wang, J. Q. Zhang, X. M. Guo,
Two kinds of contact problems in decagonal quasicrystalline matirials on point group $10mm$, Acta Mechanica Sinica. 37:169-174(2005).

[8]P. Deift, and X. Zhou, A steepest descent method for oscillatory Riemann-Hilbert problems,
Ann Math. 137,295-368(1993).

[9]J. Lenells, Initial-boundary value problems for integrable evolution equations with $3\times3$ Lax pairs, Phys D. 241:857-875(2012).

[10]J. Lenells, The Degasperis-Procesi equation on the half-line, Nonlinear Anal. 76:122-139(2013).

[11]J. Xu, and E. G. Fan, The unified method for the Sasa-Satsuma equation on the half-line, Proc R Soc A, Math Phys Eng Sci. 469:1-25(2013).

[12]J. Xu, and E. G. Fan, The three wave equation on the half-line, Phys Lett A. 378:26-33(2014).

[13]J. Xu, and E. G. Fan, Initial-boundary value problem for the two-component nonlinear Schr\"{o}dinger equation on
the half-line, Journal of Nonlinear Mathematical Physics. 23:167-189(2016).

[14]X. G. Geng, H. Liu, and J. Y. Zhu, Initial-boundary value problems for the coupled nonlinear Schr\"{o}dinger equation on the half-line, Stud Appl Math. 135: 310-346(2015).

[15]H. Liu, and X. G. Geng, Initial-boundary problems for the vector modified Korteweg-deVries equation via Fokas unified transform method, J Math Anal Appl. 440:578-596(2014).

[16]S. F. Tian, Initial-boundary value problems for the general coupled nonlinear Schr\"{o}dinger equation on the interval via the Fokas method, Journal of Differential Equations. 262:506-558(2017).

[17]M. Hisakado, T. Iizuka, and M. Wadati, Coupled Hybrid Nonlinear Schr\"{o}dinger Equation and Optical Solitons J. Phys. Soc.
Jpn. 63:2887-2894(1994).

[18]M. Hisakado and M. Wadati, Integrable Multi-Component Hybrid Nonlinear Schr\"{o}dinger Equations,
J. Phys. Soc. Jpn. 64:408-413(1995).

[19]K. Porsezian, Soliton models in resonant and nonresonant optical fibers, Pramana J. Phys. 57:1003-1039(2001).

[20]H. C. Morris, and P. K. Dodd, The Two Component Derivative Nonlinear Schr\"{o}dinger Equation,
Phys. Scr. 20:505-508(1979).

[21]L. M. Ling and Q. P. Liu, Darboux transformation for a two-component derivative nonlinear Schr\"{o}dinger equation, J. Phys. A: Math. Theor. 43:434023(2010).

[22]M. Li, B. Tian, W. J. Liu, Y. Jiang, and K. Sun,
Dark and anti-dark vector solitons of the coupled modified nonlinear Schr\"{o}dinger equations from the birefringent optical fibers, Eur. Phys. J. D. 59:279-289(2010).

[23]A. Janutka, Collisions of optical ultra-short vector pulses, J. Phys. A. 41:285204 (2008).

[24]H. Q. Zhang, B. Tian, X. L¨¹, H. Li, and X. H. Meng,
Soliton interaction in the coupled mixed derivative nonlinear Schr\"{o}dinger equations, Physics Letters A. 373:4315-4321(2009).

[25]Yoshimasa Matsuno, The N-soliton solution of a two-component modified nonlinear Schr\"{o}dinger equation, Physics Letters A. 375:3090-3094(2011).

[26]H. Q. Zhang, Darboux Transformation and N-Soliton Solution for the Coupled
Modified Nonlinear Schr\"{o}dinger Equations, Zeitschrift f\"{u}r Naturforschung A. 67(12):711-722(2012).

\end{document}